 \documentclass[showpacs,amsmath,amssymb,aps,floatfix]{revtex4}  

\usepackage{graphicx}
\usepackage{dcolumn}
\usepackage{bm}
\begin{document}  
\title{Microscopic analysis of $K^{+}$-nucleus elastic scattering based  
on $K^{+}$-nucleon phase shifts}  
\author{H. F. Arellano}  
\email{arellano@dfi.uchile.cl}  
\affiliation{Departamento de F\'{\i}sica,  
             Facultad de Ciencias F\'{\i}sicas y Matem\'aticas \\  
             Universidad de Chile, Casilla 487-3, Santiago, Chile}  
\author{H. V. von Geramb}  
\email{geramb@uni-hamburg.de}  
\affiliation{Theoretische Kernphysik, Universit\"at Hamburg \\  
             Luruper Chaussee 149, D-22761, Hamburg, Germany}  
\date{\today}

  
\begin{abstract}  
We investigate $K^{+}$-nucleus  elastic scattering at   
intermediate energies within a microscopic optical model approach.  
To this effect we use the current $K^{+}$-nucleon {\it (KN)} 
phase shifts from   the Center for Nuclear Studies   
of the George Washington University as primary input.   
First, the {\it KN} phase shifts are used to generate 
Gel'fand-Levitan-Marchenko real and local inversion potentials.   
Secondly, these potentials are supplemented with a short range   
complex separable term in such a way that the corresponding   
unitary and non-unitary {\it KN} $S$ matrices are exactly 
reproduced.  
These {\it KN} potentials allow to calculate all needed on- and off-shell   
contributions of the $t$ matrix,   
the driving effective interaction in the full-folding 
$K^{+}$-nucleus optical model potentials reported here.   
Elastic scattering of positive kaons from $^{6}$Li, $^{12}$C, $^{28}$Si   
and $^{40}$Ca are studied at beam momenta in the range 400-1000 MeV/{$c$},
leading to a fair description of most differential and total cross 
section data.
To complete the analysis the full-folding model, three kinds of    
simpler $t\rho$ calculations are considered and results discussed.  
We conclude that conventional medium effects, in conjunction with
a proper representation of the basic {\it KN} interaction are essential for the
description of $K^{+}$-nucleus phenomena.
\end{abstract}  
  
  
\maketitle               
  
  
\section{Introduction}  
Over the past two decades the study of $K^{+}$-nucleus {\it (KA)} 
collisions with light targets received considerable attention both 
experimentally and theoretically \cite{Fri97,Fri97b,Ker00}.  
This has been so mainly in virtue of the smooth energy dependence, the   
relative weak strength of the $K^{+}$-nucleon {\it (KN)} interaction  
and the strangeness of projectile.   
Herewith, it was expected and largely confirmed that intermediate energy   
$t\rho$ optical potentials would suffice to describe the scattering data.  
However, some unexpected and persistent shortcomings were observed in the   
description of total cross section data, taken  in transmission experiments   
at beam momenta in the range 500-1000 MeV/$c$ \cite{Kra92,Fri97,Fri97b}.  
This situation has triggered outlook  for new    
physics with models including unconventional   
as well as higher order effects \cite{Mard90,Che99}.  
As an important and unsatisfactory  element, in all these discussions  
and up to now, remains the absolute normalization error of the measured cross   
sections, being  
$\pm17$\% \cite{Ker00}. To mention a few of these efforts, {\em viz}     
covariant formulation \cite{Ker00,Che99},   
consideration of  medium modifications of the {\it KN} interaction  
within the target nucleus environment \cite{Cai95,Cai96}, 
the use of on- and off-shell $t$-matrix contributions  
with the construction of separable scattering amplitudes \cite{Jia95},
and the possible manifestation of $\Theta^{+}$ pentaquark in {\it KA}
collissions \cite{Gal05}.  
  
The microscopic optical model potential (OMP) approach we present here  
embodies most of the above elements but puts emphasis on the best possible    
direct use of {\it KN} phase shift data to the generate on- and off-shell 
{\it KN} $t$-matrix elements.   
This is achieved with the construction of a {\it KN} potential,   
in true sense a {\it KN} optical model potential when the respective $S$ 
matrix is non unitary, which reproduces in absolute terms the phase shift data.  
This approach distinguishes several steps.   
First, for each partial wave and set of {\it KN} data,   
an optimal Gel'fand-Levitan-Marchenko inversion potential   
$V_{\alpha}(r)$ is calculated \cite{San97,Fun01}.   
Second, a short range rank-one separable potential,  
with energy dependent and possibly complex strengths,   
is added to $V_\alpha(r)$ and matched to the data \cite{Fun01}.    
These potentials are used in Lippmann-Schwinger equations for the   
{\it KN} $t$ matrix,   
the defined effective interaction in the full-folding optical  
model approach discussed here.    
These folding calculations are carried out in momentum space with the   
use of nonlocal single particle target densities.   
Herein, the full-folding calculations use only the {\it KN} $t$ matrix and   
thus neglect the Pauli blocking in the propagation of nucleons in the   
nucleus \cite{Are02}.

This article is organized as follows.  
In Section II we summarize the main relativistic considerations in the 
definition of the bare {\it KN} potential. We also outline some aspects of 
quantum inverse scattering relevant to our applications, specify the {\it KN} 
data and  discuss the implied features in the {\it KN} interaction.    
In Section III we present the salient features of the full-folding {\it KA}    
optical potential and discuss three alternative $t\rho$ approximations.   
In Section IV we show and discuss {\it KA} elastic scattering applications  
for selected nuclei.  
In section V we present a summary and conclusions of this analysis.  
  
  
\section{The {\it KN} effective interaction}  
We base our study on  the current {\it KN}  partial wave phase shift   
solutions, single and continuous energy solutions for
$0 < T_\textrm{Lab} < 1$GeV,   
of Richard Arndt {\em et al.} and retrieved data from    
Center for Nuclear Studies (CNS)   
of the George Washington University (GWU) \cite{SAID,Hyl92}.  
These data are sufficient to specify the partial wave $S$ matrix   
or $t$ matrix on shell.   
However, these quantities alone are insufficient in the context of  
the many-body approach since the {\it KA} optical model requires  
the $t$ matrix off shell.  
Thus, the problem is ill posed and requires a theoretical extension of the  
on-shell $t$-matrix into the off-shell domain.   
The solution to this problem is not unique.   
However, since our analysis hinges upon a potential theory   
we chose  a {\it KN} potential concept also for this purpose.   
The off-shell extension of the $t$-matrix interaction takes into account  
free particle propagation, counted to all orders in the ladder  
approximation, but without {\it KA} medium effects. 
We calculate the  $t$ matrices with a 
Lippmann-Schwinger equation in momentum space.  
         
A well established and often applied link between phase shifts and   
potential is the inverse scattering formalism of Gel'fand-Levitan   
and Marchenko. The present application points towards use of
the fixed angular momentum or partial wave 
Schr\"odinger-type equation version, mathematically speaking 
Sturm-Liouville equation, yielding
energy independent and local potentials~\cite{San97}. 
Before we enter into the more technical aspects of inversion it 
appears useful to recall the relativistic aspects which we associate 
with the relative motion Schr\"odinger-type equation  
being used for the {\it KN} pair in the C.M. system. 

To formulate the relativistic Hamiltonian description of quantum mechanical
particle sytems, the standard rules 
for constructing the momenta from the Lagrangian
cannot be applied when the Lagrangian is singular and the set $(q,\dot{q})$
and $(q,p)$ are not uniquely related.  A Lagrangian is singular
when the Hessian vanishes, 
\begin{equation}
det\left(\frac{\partial^2 L}{\partial \dot{q}_i \partial \dot{p}_j}\right)=0.
\end{equation}
In such a case it is not possible to simply eliminate the velocity
dependences by  coordinates and momenta. In order to obtain a Hamiltonian, 
which depends on coordinates and momenta only, it requires
additional independent constraining relations $\phi(q,p)_j\approx 0$ to
achive uniqueness.
Dirac gave very general rules to construct the Hamiltonian and calculated 
sensible brackets that can be used to describe classical and 
quantum mechanical dynamics~\cite{dir64}.
These rules are generally known as Dirac's constraint dynamics. 
Out of several, an instant form dynamics, developed and  used by Crater and 
Van Alstine~\cite{cra90,liu03},
is used herein to foster the relativistic nature of the
partial wave equation which we use in phase shift inversion. 

The Hamiltonian is obtained by a Legendre transformation of the Lagrangian, 
where  all  primary constraints are multiplied by arbitrary 
functions of time $\lambda_j$ and  added to $H$. This yields the 
total Hamiltonian ${\mathcal H}$,
\begin{equation}
{\mathcal H}=H+\sum_{j=1}^M\lambda_j\phi_j(q,p).
\end{equation}
For consistency, the constraints must not change under a time evolution 
\begin{equation}
\dot\phi_j=[\phi_j,{\mathcal H}]\approx 0.
\end{equation}
Therefrom  three cases are distinguised: an equation 
can give an identity, it can give a linear equation for the $\lambda_j$ or 
it  gives an equation containing selectively $p$ and $q$, in which case it 
must be considered as another constraint. The constraints that arise from
this procedure are called secondary. Also, any linear combination of 
constraints is again a constraint.

Crater {\em et al.}~\cite{cra90,liu03} studied  the relativistic  2-particle 
dynamics as a problem  with constraints of the kind
\begin{equation}\label{tod1}
\phi_i(q,p):= m_i^2+p^2_i+\Phi_i(q,p,s),\quad i=1,2,
\end{equation}
where, in view of the final result, we adopted their metric i.e.    
using the other sign in the four product, $g_{\mu\mu}=
(-1,1,1,1)$. The two constraints are limited to yield
$\phi_i(q,p)\approx 0$,
being  generalized mass shell constraints for the two particles. 
\begin{equation}
\mathcal{H}_{i}=p_{i}^{2}+m_{i}^{2}+\Phi_{i}(x,p_{1},p_{2})\approx 0,
\quad i=1,2,
\end{equation}
where $(\mathcal{H}_{1},\mathcal{H}_{2})$ are covariant constraints 
on the four momenta $(p_1,p_2)$. 
The interaction functions $(\Phi_1,\Phi_2)$  being equal, 
$\Phi_1=\Phi_2$, implies a relativistic analog of Newton's third 
law with relative distance  $x=x_{1}-x_{2}$.
The total Hamiltonian $\mathcal{H}$ from these constraints alone is
\begin{equation}
\mathcal{H=\lambda}_{1}\mathcal{H}_{1}+\lambda_{2}\mathcal{H}_{2}, 
\end{equation}
containing the two Lagrange multipliers $\lambda_{i}$.
Thus the quantum mechanical particle constraints become  
Schr\"odinger type equation
$\mathcal{H}_{i}|\psi\rangle=0$.
In order that each of these constraints be conserved, the C.M. eigentime $\tau$
is used for the temporal evolution
\begin{equation}
\lbrack\mathcal{H}_{i},\mathcal{H}]|\psi\rangle=
i\frac{d\mathcal{H}_{i}}{d\tau}|\psi\rangle=0.
\end{equation}
Thus
\begin{equation}
\{[\mathcal{H}_{i},\lambda_{1}]\mathcal{H}_{1}|\psi\rangle+\lambda
_{1}[\mathcal{H}_{i},\mathcal{H}_{1}]|\psi\rangle+[\mathcal{H}_{i},\lambda
_{2}]\mathcal{H}_{2}|\psi\rangle+\lambda_{2}[\mathcal{H}_{i},\mathcal{H}%
_{2}]\}|\psi\rangle=0,
\end{equation}
which equals  the compatibility condition
\begin{equation}\label{compateqn}
\lbrack\mathcal{H}_{1},\mathcal{H}_{2}]|\psi\rangle=0.
\end{equation}
This condition guarantees that, with the Dirac Hamiltonian $\mathcal{H}$,
the system evolves such that the motion is constraint to
the surfaces on the mass shells described by the constraints  
$\mathcal{H}_{1}$ and $\mathcal{H}_{2}$. 
Eq. (\ref{compateqn}) constraints the interaction
\begin{equation}
[p_1^2,\Phi_2]-[p_2^2,\Phi_1]+[\Phi_1,\Phi_2]\approx 0,
\end{equation}
with the sensible solution  
$ \Phi_{1}=\Phi_{2}=\Phi(x_{\perp},p_1,p_2)$.
The transverse coordinate  
$x_{\nu\perp}:=(g_{\nu\mu}-P_{\nu}P_{\mu}/P^{2})x^\mu$ is used. 
A brief summary of all the used relevant quantities, 
the so called Todorov variables, makes it easy to
follow the last few steps: 
total momentum $P=p_1+p_2$,
total C.M. energy $w = \sqrt{-P^2}$,
relative position $x:=x_{1}-x_{2}$,
relative momentum $p:=(\epsilon_2p_1-\epsilon_1p_2)/w$ with
C.M. single particle energies  
  $\epsilon_1=(w^2+m_1^2-m_2^2)/2w$ and  
  $\epsilon_2=(w^2+m_2^2-m_1^2)/2w$,  $\hat{P}=P/w$, 
$p_1=\epsilon_1\hat{P} +p$ and $p_2=\epsilon_1\hat{P}-p$. 
As generalization of the nonrelativistic
reduced mass $\mu=m_1m_2/(m_1+m_2)$ we are to distinguish two quantities: 
the relativistic reduced mass $ m_w:=(m_1m_2)/w$, and the
reduced energy  $\epsilon_w:=(w^2-m_1^2-m_2^2)/2w$.  
The on-shell relative momentum is given by
$b^2(w)=\epsilon_w^2-m_w^2=\epsilon_1^2-m_1^2=\epsilon_2^2-m_2^2$,
or the wave number $k^2=b^2(w)$.

To define the relative momentum, we require  the difference 
$\mathcal{H}_{1}-\mathcal{H}_{2}\approx 0$ independent of the
interaction $\Phi$. Thus 
\begin{equation}
\mathcal{H}_{1}-\mathcal{H}_{2}=m_1^2+p_1^2+
\Phi-m_2^2-p_2^2-\Phi\approx 0, 
\end{equation}
which implies
\begin{equation}
 2P\cdot p+(\epsilon_2-\epsilon_1)w+m_1^2-m^2_2\approx 0.
\end{equation}
In the C.M.  frame:  $P\cdot p=0$ and  
$P=(-P^0,{\bf P}=0)$ has the implication that the relative momentum
$p=(0,{\bf p})$ has a vanishing time like component.
From
\begin{align}
p_1=(\epsilon_1,+{\bf p}),&\quad  p^2_1=-\epsilon_1^2+
{\bf p}^2=-m_1^2 \nonumber \\
p_2=(\epsilon_2,-{\bf p}),&\quad  p^2_2=-\epsilon_2^2+
{\bf p}^2=-m_2^2,
\end{align}
follows simply $P=p_1+p_2=-(\epsilon_1+\epsilon_2)$.
The vanishing time like relative momentum implies
\begin{align}
\epsilon_{1}  &  =\frac{w^{2}+m_{1}^{2}-m_{2}^{2}}{2w},\nonumber \\
\epsilon_{2}  &  =\frac{w^{2}+m_{2}^{2}-m_{1}^{2}}{2w}, 
\end{align}
and 
\begin{align}
\mathcal{H}_{1}&={\bf p}^2+\Phi-\epsilon_1^2+m_1^2,\nonumber \\
\mathcal{H}_{2}&={\bf p}^2+\Phi-\epsilon_2^2+m_2^2.
\end{align}
The  combination yields  a stationary Schr\"odinger type  wave
equation for an effective C.M. single particle with mass $m_w$ and energy $\epsilon_w$
\begin{equation}
\mathcal{H}|\psi\rangle=
\frac{(\epsilon_2\mathcal{H}_{1}+\epsilon_1\mathcal{H}_{2})}{w}|\psi\rangle=
({\bf p}^2+\Phi-b^2)|\psi\rangle = 0.
\end{equation}
In the C.M. system the relative energy and time are removed from the
problem, $p=p_{\perp}=(0,{\bf p})$ and
$x_{\perp}=(0,{\bf r})$ implies  
\begin{equation}
\label{kleing}
\left( {\bf p}^2+\Phi({\bf r})-b^{2}\right)  |\psi\rangle=0.
\end{equation}
The relation to the stationary non relativistic Schr\"odinger 
equation is herewith established. It requires only 
spherical coordinates and a partial wave expansion to have  
the radial type Sturm Liouville equations
for Gel'fand-Levitan-Marchenko fixed angular momentum inversion.

The inversion  method is useful and physically justified  for cases in which the   
phase shifts are smooth functions of the energy and resonances are absent.   
However, actual {\it KN} phase shift data are not perfectly smooth,   
show significant error bars and are not free of personal preferences.   
To keep these preferences at a minimum  we divide the {\it KN} partial wave   
potentials into two parts.   
The first part is the result of a Gel'fand-Levitan-Marchenko inversion   
with an optimal smooth rational function  fit to the unitary $S$-matrix   
sector of the data.   
The resulting real potentials are smooth functions of the radial distance  
and play the role of what we call a {\em reference potential}.
  
The basic equations of inversion are  
the  radial Schr\"odinger equation  
\begin{equation}  
\left[ -\frac{d^2}{ dr^2} + \frac{\ell ( \ell+1)}{r^2 } +  
2 \mu V_\ell (r)  
\right] \psi_\ell (k,r) = k^2 \psi_\ell (k,r),  
\qquad 0\le r < \infty  
\label{rse}  
\end{equation}  
where $V_\ell (r)$ is a local  energy independent but explicitly 
partial wave dependent   
coordinate space potential. The factor $2\mu$ is used   
to make a comparison with nonrelativistic potentials more obvious.  
The right hand side refers to the relative two-particle momentum  
or wave number $k$ which is related to the kinetic energy of the   
kaon in the laboratory system $T_\textrm{Lab}$, its mass $m_K$ and the   
nucleon mass $m_N$, by means of  
\begin{equation}  
      s=(m_N+m_K)^2+2m_NT_\textrm{Lab}  
\end{equation}  
and   
\begin{equation}  
      k^2=\frac{s^2+(m_K^2-m_N^2)^2- 2 s(m_K^2+m_N^2)}{4 s }.  
\end{equation}  
The boundary conditions for the physical solutions are  
\begin{equation}   
\lim_{r\to 0} \psi_\ell(k,r)=0  
\end{equation}  
and  
\begin{equation} \lim_{r \rightarrow \infty} \psi_\ell (k,r) =  
\exp [ i {\delta_\ell (k)}] \sin [ kr - \frac{\ell \pi }{ 2}  
+ {\delta_\ell (k)}]  
\end{equation}  
  
The Gel'fand-Levitan and Marchenko inversion are two different algorithm   
which should yield the exactly the same potential results. The use and comparison of  
both calculations guarantees robust results.  
  
The experimental information enters in the Marchenko inversion {\em via}   
the partial wave $S$ matrix, which is related to the scattering phase shifts   
by the relation  
\begin{equation}  
\label{sm}  
S_{\ell}(k)=\exp[2i\delta_{\ell}(k)].  
\end{equation}  
We use a rational function interpolation and  extrapolation of  
real data $\delta_\ell(k)$,    
\begin{equation} \label{delrat}  
\delta_\ell (k) = \sum_{m=1}^M \frac{D_m }{ k - d_m}  
\end{equation}  
with the asymptotic conditions  
\begin{equation}  
\lim_{k\to0} \delta_\ell (k) \sim k^{2 \ell + 1} \qquad  
\mbox{and} \qquad   
\lim_{k\to\infty} \delta_\ell (k) \sim k^{-1}.  
\end{equation}  
In any case there are few poles $d_m$ and strengths $D_m$  sufficient   
to provide a smooth description of data. For the {\it KN} system, there are no
bound state to be extracted, thus we simply    
use a rational function interpolation and extrapolation of real data
$\delta_\ell(k)$ with a fully symmetric distribution of poles and
zeros in the upper and lower half k-plane. This implies that the boundary
conditions at the origin and the infinity  are  satisfied. Furthermore,
using a symmetric Pad\'e approximant for the exponential function
guarantees that the number of zeros and poles of the S-Matrix, in the
upper and lower half complex k-plane,
are the same, the index is zero and no bounds are present. 

Using a $[4/4]$  Pad\'{e} approximation   
for the exponential function $e^z$ is highly accurate and  
substituting the rational phase function    
into $z = 2i\delta_\ell(k)$ gives a rational $S$ matrix   
\begin{equation} \label{smrat}  
S_\ell (k) = 1 + \sum_{n=1}^{2N} \frac{s_n }{ k - \sigma_n}  
= \prod_{n=1}^{N} \frac{k + \sigma_{n}^\uparrow }{ k - \sigma_{n}^\uparrow}  
\cdot \frac{k + \sigma_{n}^\downarrow }{ k - \sigma_{n}^\downarrow},  
\end{equation}  
where we denote  
$\{ \sigma_n^\uparrow \} := \{ \sigma_n | \mbox{Im} (\sigma_n) > 0 \}$  
and $\{ \sigma_n^\downarrow \} := \{ \sigma_n | \mbox{Im} (\sigma_n) < 0 \}.$  
The Marchenko input kernel   
\begin{equation} \label{minpk}  
F_\ell (r,t) = -\frac{1}{2\pi} \int_{-\infty}^{+\infty} h^+_\ell(kr)    
  \left[ S_\ell(k)-1 \right] h^+_\ell(kt) dk  
\end{equation}  
is readily computed using  
Riccati-Hankel functions $h^+_\ell(x)$ and contour integration.  
This implies an algebraic equation for the translation kernel $A_\ell(r,t)$  
of the Marchenko equation  
\begin{equation} \label{mfeqn}  
A_\ell (r,t)+F_\ell (r,t)+\int_{r}^{\infty}A_\ell(r,s)F_\ell(s,t)ds = 0.  
\end{equation}  
The potential is obtained from the translation kernel derivative  
\begin{equation}  
\label{potm}  
V_{\ell}(r)=-2\frac{d}{dr}A_{\ell}(r,r).  
\end{equation}  
Thus, the rational representation of the scattering data leads to  
an algebraic form of the potential.  
  
The Gel'fand-Levitan inversion uses Jost functions as input.   
The latter is related to the $S$ matrix by  
\begin{equation}  
\label{rhp}  
S_{\ell}(k)=\frac{F_{\ell}(-k)}{F_{\ell}(k)}.  
\end{equation}  
Using the representation (\ref{smrat}), the Jost function in   
rational representation is given by  
\begin{equation} \label{jost_1}  
F_\ell (k) = \prod_{n=1}^{N}   
\frac{k - \sigma_{n}^\downarrow }{k + \sigma_{n}^\uparrow}  
=  
 1 + \sum_{n=1}^N \frac{B_n }{k + \sigma_n^\uparrow},  
\end{equation}  
or  
\begin{equation}  
|F_\ell (k)|^{-2} = 1 + \sum_{n=1}^{N}   
\frac{ L_n }{ k^2 - \sigma_{n}^{\downarrow 2}}.  
\end{equation}  
The input kernel  
\begin{equation}  
G_\ell (r,t) = \frac{2}{\pi} \int_{0}^{\infty} j_\ell (kr) \left[  
             \frac{1}{|F_\ell (k)|^2}- 1  
 \right] j_\ell (kt) dk,  
\end{equation}  
where $j_{\ell}(x)$ represent the Riccati-Bessel functions, is analytic.  
The Gel'fand-Levitan equation  
\begin{equation}  
K_\ell (r,t)+G_\ell (r,t)+\int_{0}^{r}K_\ell (r,s)G_\ell (s,t)ds = 0,  
\end{equation}  
relates input and translation kernels, where the potential is defined by  
\begin{equation}  
\label{vgl}  
V_{\ell}(r)=2\frac{d}{dr}K_{\ell}(r,r).  
\end{equation}  
Thus, also this potential has an algebraic form.  
  
The second part is a short-range rank-one separable potential         
with  real or complex energy-dependent strengths fixed to the actual data.  
This idea has been developed and implemented   
in nucleon-nucleon studies and applied to nucleon-nucleus   
scattering \cite{Fun01,Are02}.   
Here, we use the {\it KN} potential as the sum of a local inversion  
potential $V_\alpha(r)$ supplemented with a  separable term   
\begin{equation} \label{VKNrspace}  
V_{KN}(r,r',E) = V_\alpha(r')\frac{\delta(r-r')}{rr'}+  
\phi_\alpha(r) \Gamma_\alpha(E)\phi_\alpha(r').  
\end{equation}  
The partial waves are identified with $\alpha$ and  
$\Gamma_\alpha(E)$ are energy-dependent strengths with imaginary component  
for those channels where the $S$ matrix is not unitary.   
This is the case of only some partial wave data.  
For a given reference potential $V_\alpha(r)$ and data, the determination   
of $\Gamma_\alpha(E)$ is a straightforward procedure \cite{Fun01}.  
  
\begin{figure}[!hbt]  
\includegraphics[scale=0.50,angle=-90]{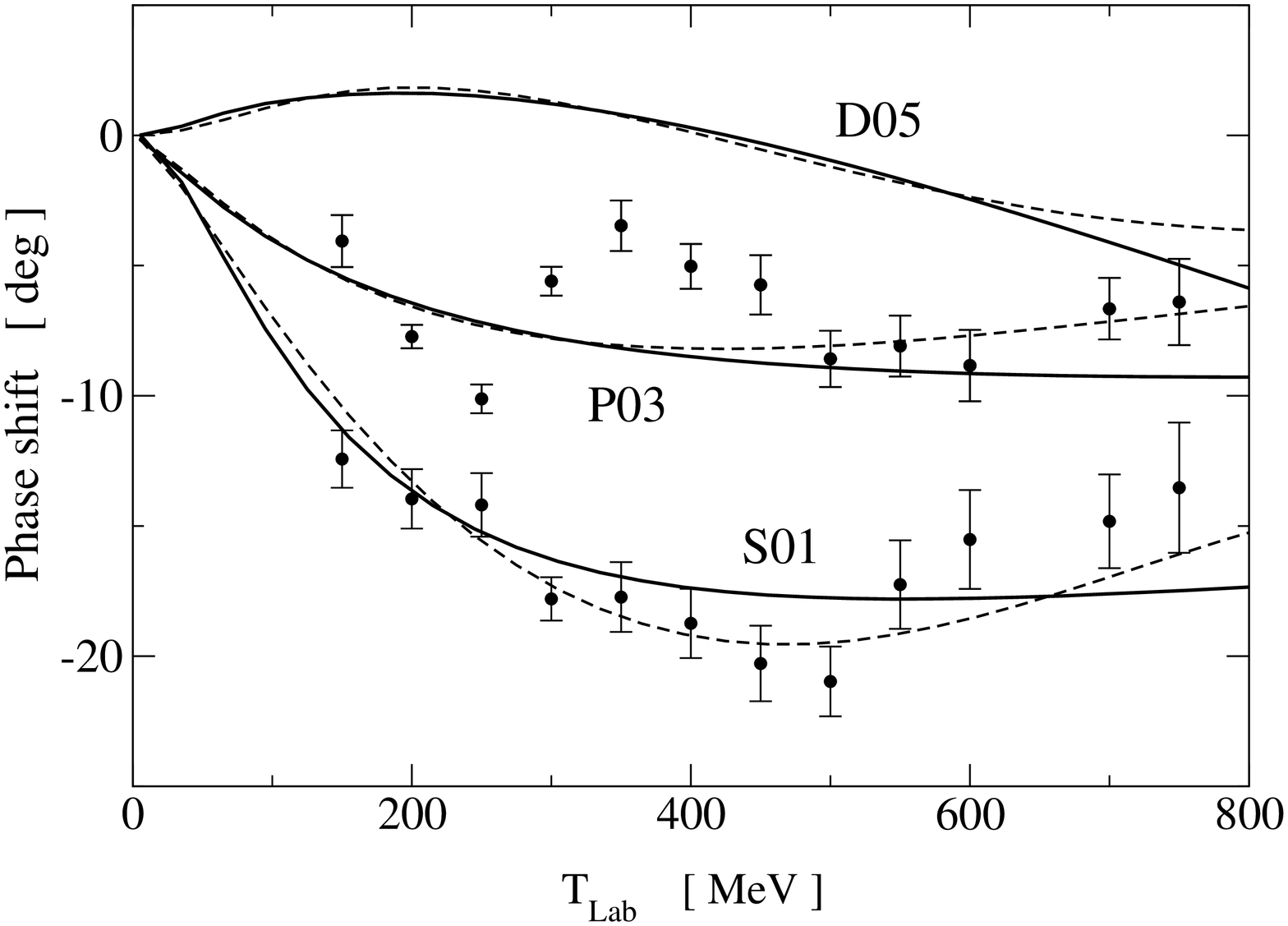}  
\vspace{-5mm}  
\caption{{\protect\small  
\label{stretched_0}  
        The isospin-zero stretched states phase shifts as function of the  
        $K^{+}$ kinetic energy.   
        The single- and continuous-energy solutions of the GWU   
        analyses are represented with large and small circles.  
        The solid curves represent the phase shifts from the reference   
        potential.  
        }}  
\end{figure}

\begin{figure}[!hbt]  
\includegraphics[scale=0.50,angle=-90]{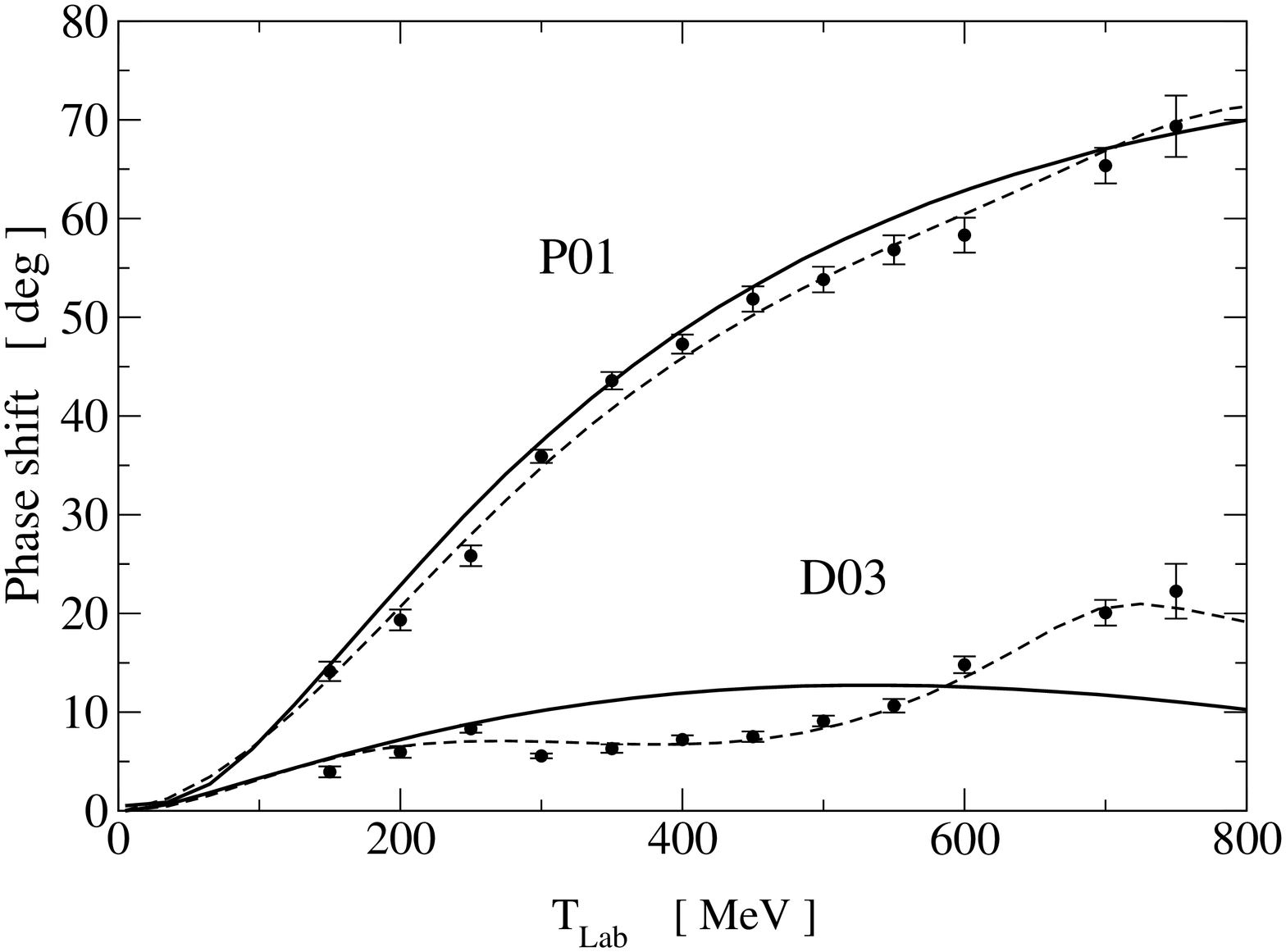}  
\vspace{-5mm}  
\caption{{\protect\small  
\label{unstretched_0}  
        The same as Fig. \ref{stretched_0} but for the unstretched states.  
        }}  
\end{figure}

\begin{figure}[!hbt]  
\includegraphics[scale=0.50,angle=-90]{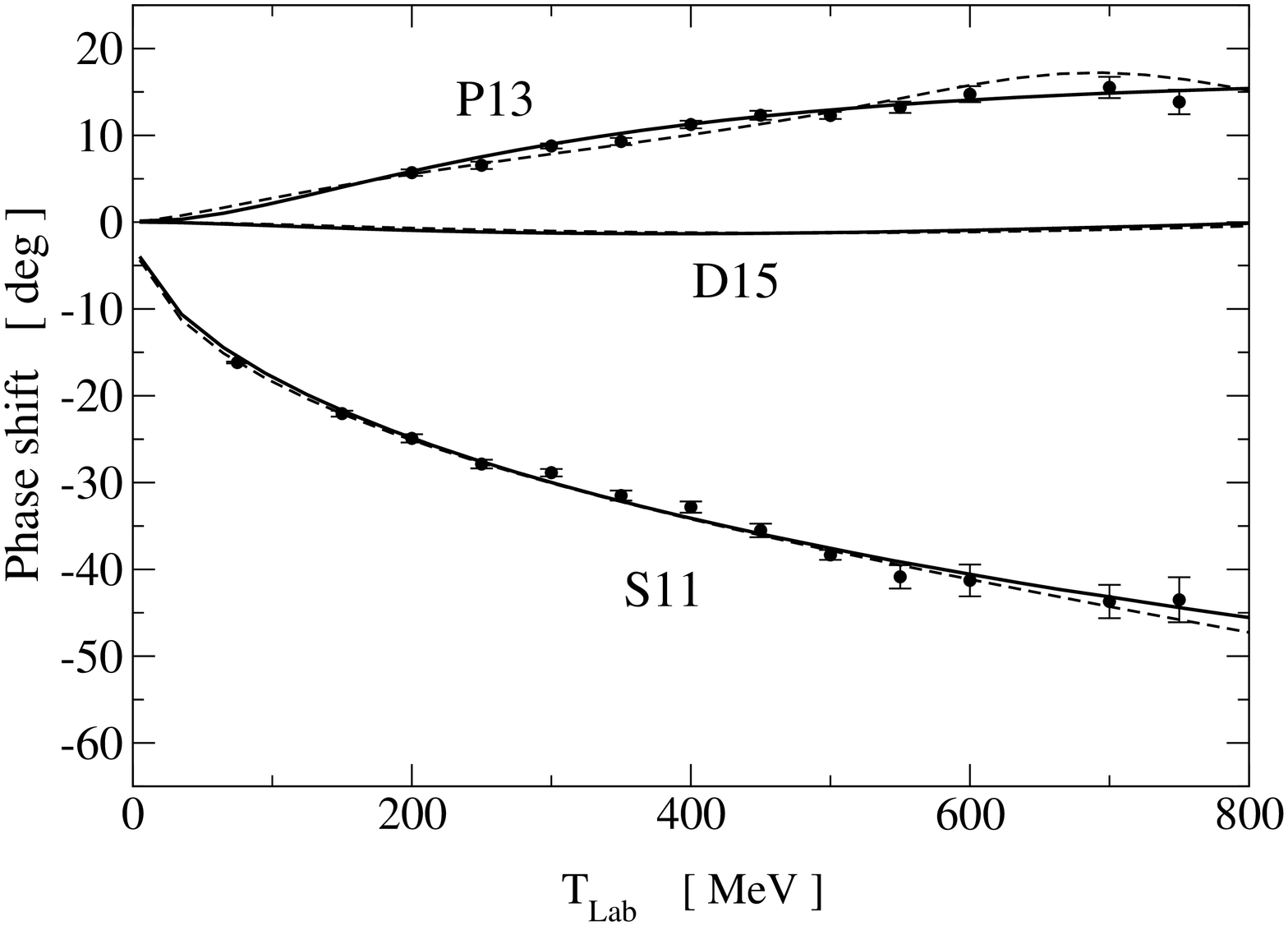}  
\vspace{-5mm}  
\caption{{\protect\small  
\label{stretched_1}  
        The isospin-one stretched states phase shifts as function of the  
        $K^{+}$ kinetic energy.   
        The single- and continuous-energy solutions of the GWU   
        analyses are represented with large and small circles.  
        The solid curves represent the phase shifts from the reference   
        potential.  
        }}  
\end{figure}

\begin{figure}[!hbt]  
\includegraphics[scale=0.50,angle=-90]{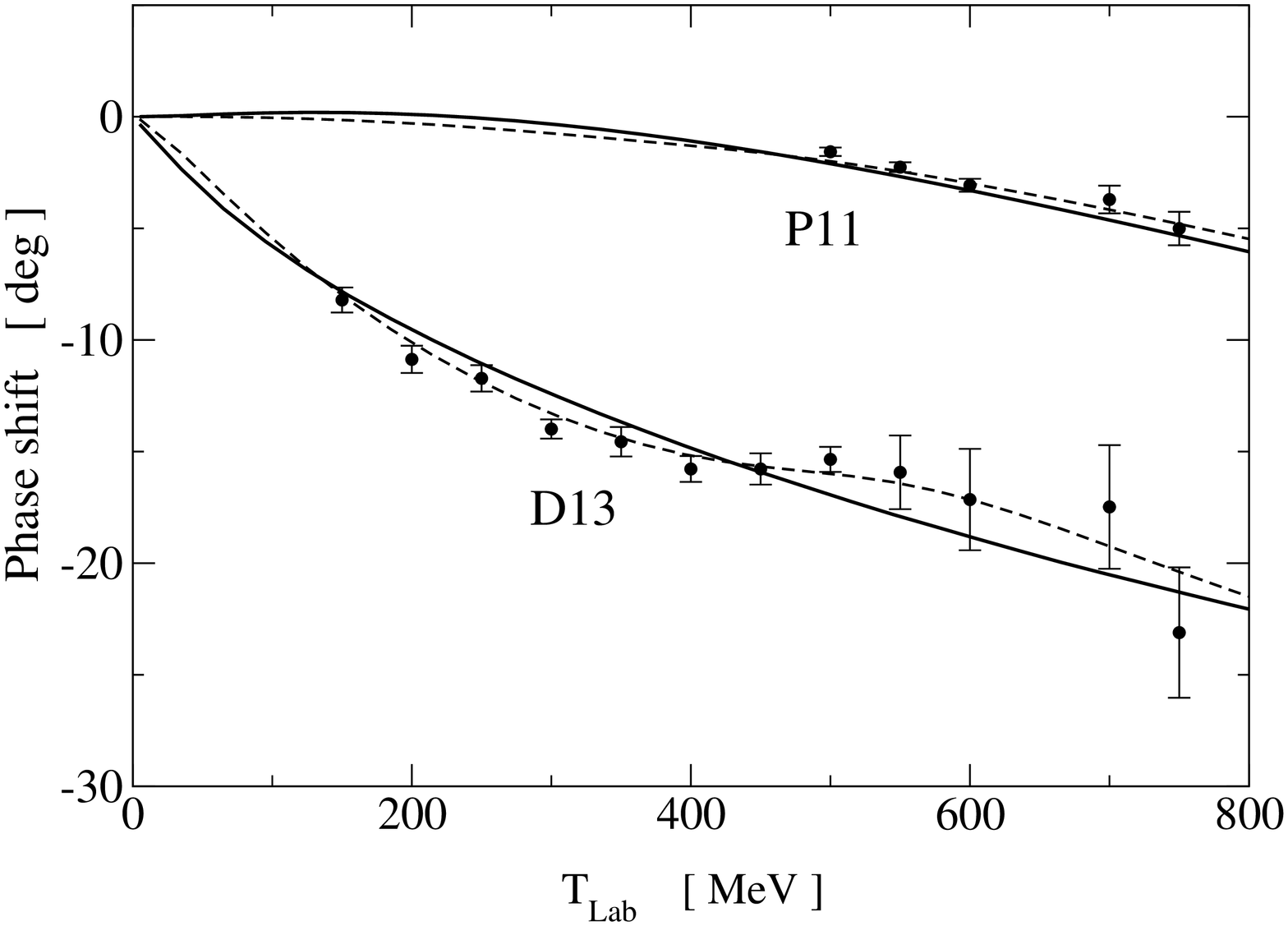}  
\vspace{-5mm}  
\caption{{\protect\small  
\label{unstretched_1}  
        The same as Fig. \ref{stretched_1} but for the unstretched states.  
        }}  
\end{figure}

\begin{figure}[!hbt]  
\includegraphics[scale=0.50,angle=-90]{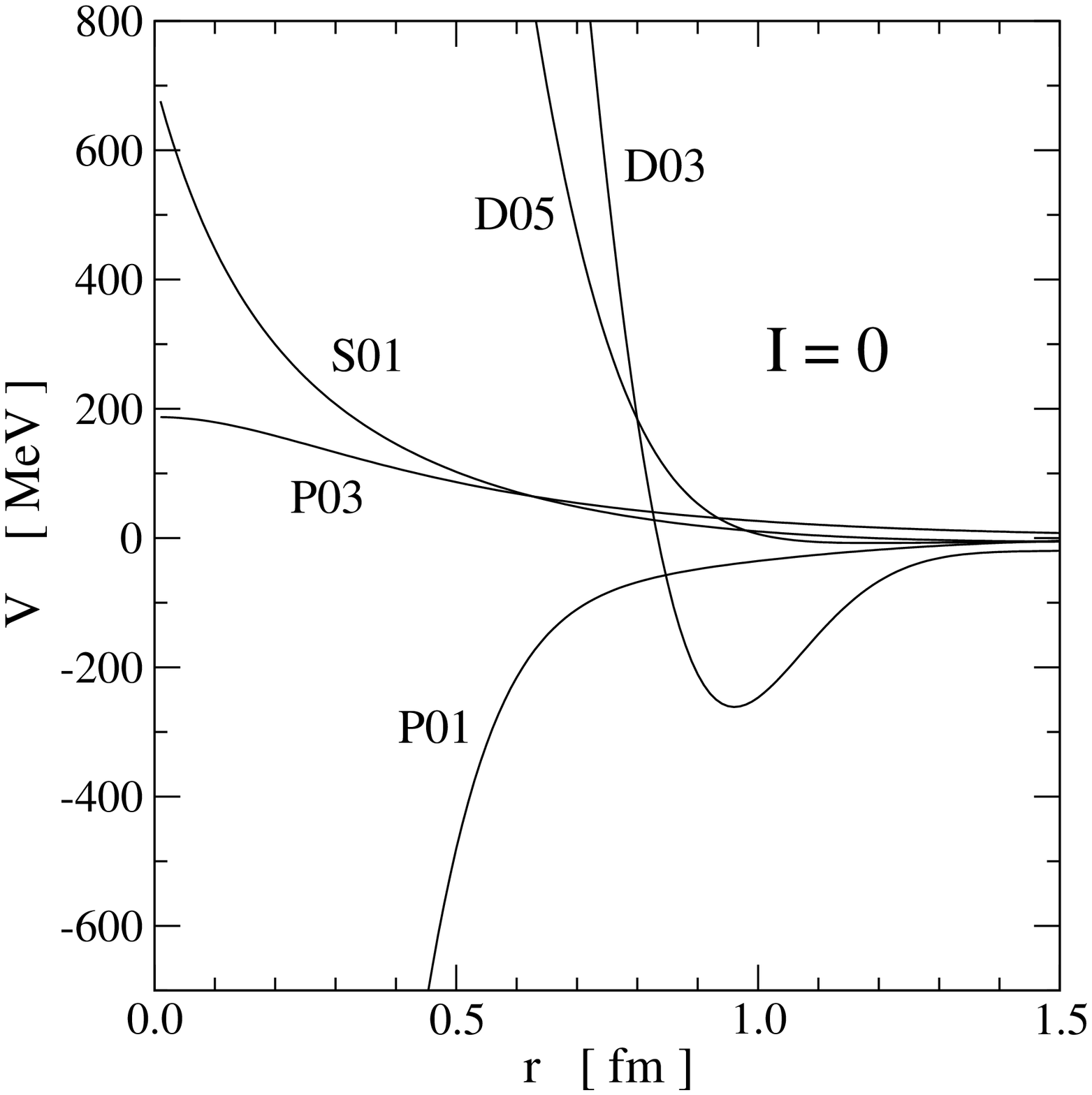}  
\vspace{-5mm}  
\caption{{\protect\small  
\label{VI0}  
       The radial dependence of the isospin-zero $K^{+}N$ reference potentials.
        }}  
\end{figure}

\begin{figure}[!hbt]  
\includegraphics[scale=0.50,angle=-90]{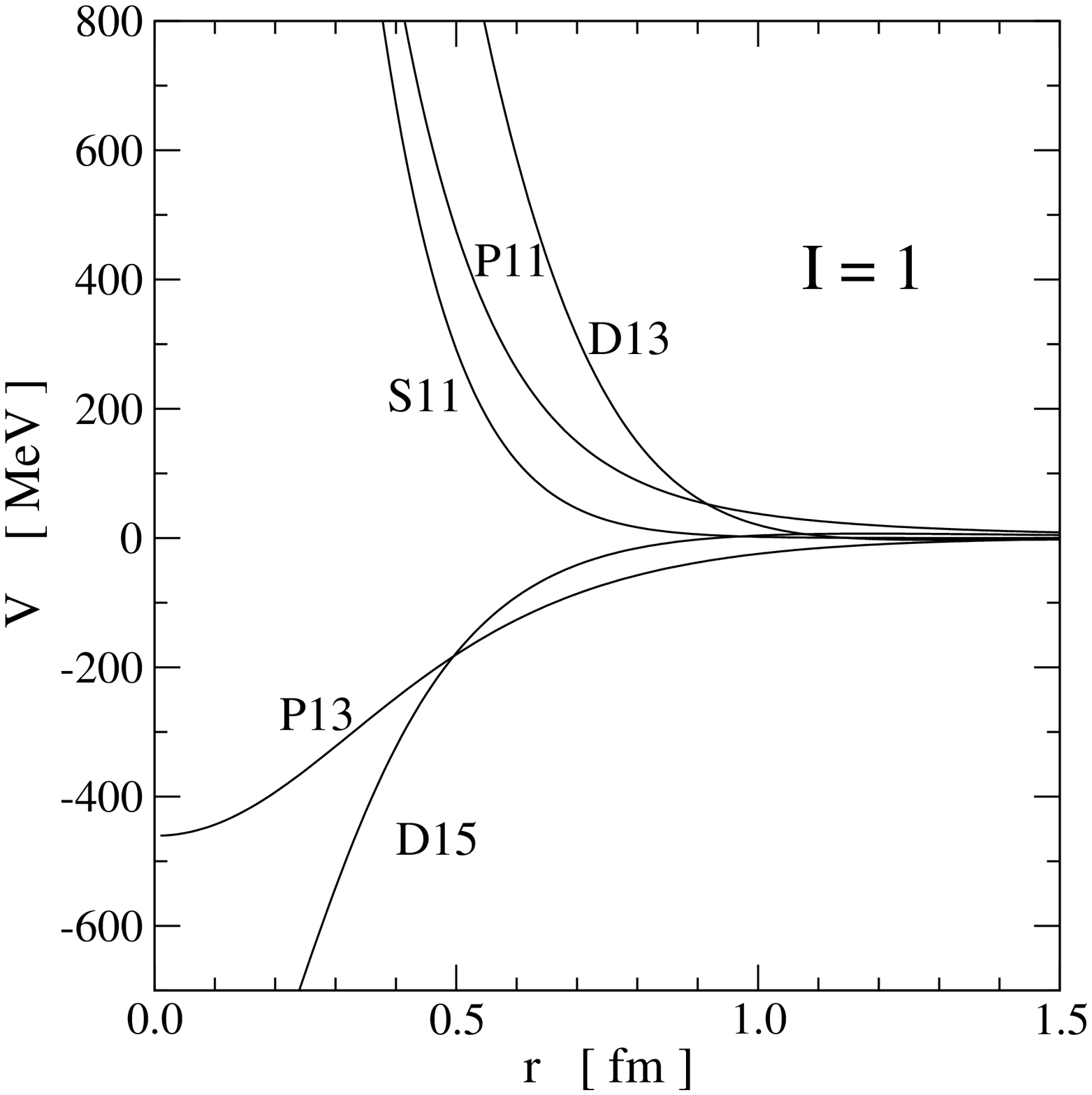}  
\vspace{-5mm}  
\caption{{\protect\small  
\label{VI1}  
       The radial dependence of the isospin-one $K^{+}N$ reference potentials.  
        }}  
\end{figure}

Thus, we base the $V_{KN}$  on the current solution of   
CNS/GWU-KN solutions \cite{SAID,Hyl92}.  
All used phase shifts, $L\leq 2$, are shown in Figs. 1-4,  
where we distinguish different data: single (full circles with error bars)  
and continuous energy (dashed curves) solutions, respectively,   
and the inversion reference potential phase shifts (solid curves)  
which reproduce the rational functions of the kinetic energy $T_\textrm{Lab}$. 
The isospin zero (I=0)  stretched ($J=L+1/2$) and anti-stretched ($J=L-1/2$)  
channels are shown in Figs.  \ref{stretched_0}   
and \ref{unstretched_0}, respectively.  
Similarly, the isospin one (I=1) stretched   
and anti-stretched channels are presented in Figs. \ref{stretched_1}  
and \ref{unstretched_1}, respectively.   
The corresponding inversion reference potentials are shown   
in Figs. \ref{VI0} and \ref{VI1}.  
In these figures  we observe that all potentials are short ranged with   
significant strengths limited to $r<1$\,fm.   
The very short range behavior depends on the  
high energy extrapolation of the rational function,  
which we did as sensible as possible.  
  
The separable potential functions are motivated and tuned to a short   
range zone in which resonances, inelastic scattering  and reactions are  
supposed to occur \cite{Fun01}, {\em viz}    
\begin{equation}  
\label{ffactor}  
\phi_\alpha(r)=N_\alpha r^{L}\exp{[-(r-r_0)^2/a^2]},  
\end{equation}  
where we have used $r_0=0.5$\, fm, $a=0.2$\, fm, with  
$N_\alpha$ a normalization constant.  
  
Any identification of resonances and reaction channels is not part   
of this endeavor.  
Thus, the separable term strengths $\Gamma_\alpha(E)$  
are fixed to the continuous energy solution partial wave phase shifts    
\cite{SAID}, whose real phase shifts are shown as dashed curves  
in Figs. 1 to 4. Vanishing imaginary phase shifts are limited to the  
channels S01, P03, D05 and F07.

  
\section{The {\it KA} optical potential}  
An optical model potential (OMP) represents an effective single-particle   
interaction potential for a projectile caused by the interaction   
with target nucleons.   
The underlying many-body problem in Brueckner's   
many-body theory yields an OMP in the form of a convolution of a   
projectile-nucleon effective interaction, the reaction matrix,   
with the target mixed density.  
  
There are many ways to obtain in practical terms a successful representation   
of effective interaction and its accurate use in the convolution integral.  
Here we use the {\it KN} $t$-matrix operator, on and off shell, as the  
effective interaction.  
Such construction has successfully been used in the past and we   
recall only its salient features to make the discussion of various   
results comprehensible.  
In the projectile-nucleus center of momentum (C.M.) reference frame,   
the collision of a projectile of kinetic energy $E$ is described by   
the full-folding OMP, which in a momentum representation   
is given by \cite{Are02}  
  
\begin{equation}  
\label{FF}  
U({\bf k}^\prime, {\bf k}; E) =   
\sum_{N=p,n}   
\int d{\bf P}\;  
\rho_{N}( {\bf P}+\frac{\bf q}{2}\; ,{\bf P}-\frac{\bf q}{2}) \;  
t_{NK^+} ( {\bf k_r}', {\bf k_r};{\bf K}+{\bf P};{s} )\;,  
\end{equation}  
where we define the mean momentum ${\bf K}=( {\bf k}^\prime + {\bf k})/2$   
and momentum transfer ${\bf q}={\bf k}^\prime - {\bf k}$.  
Here the effective interaction, in the form of the free scattering $t$ matrix, 
exhibits an explicit dependence on   
the relative momenta ${\bf k_r}$ and ${\bf k_r}'$,   
the total pair momentum ${\bf Q}={\bf K}+{\bf P}$ and   
the pair $s$ invariant.  
In particular, the relative momenta take the general form  
\begin{equation}  
\label{rel_mtum}  
{\bf k_r} = W {\bf k} - (1-W){\bf p}\; , \qquad   
{\bf k_r}^\prime = W^\prime {\bf k}^\prime-(1-W^\prime){\bf p}^\prime\; ,   
\end{equation}  
where $W$ and $W^\prime$ are scalar functions of the momenta of the colliding   
particles with relativistic kinematics built in \cite{Are02}.  
  
The momentum integral $\int d{\bf P}$ signals the folding integral.  
The intricate dependence of the many vector valued momenta makes the  
convolution quite complicated and thus the name full-folding approach  
was coined in order to signal use of the full expression, as compared to  
much simpler approximated expressions.  
Physically,  
the folding integral accounts for dynamical effects due to the Fermi motion  
as modulated by the shape of target mixed density.  
For practical reasons we represent the mixed density in terms of the local   
density $\rho(R)$ {\em via} the Slater   
approximation \cite{Are90b}, i.e.  
\begin{equation}  
\label{Slater}  
\rho( {\bf P}+\frac{\bf q}{2}\; ,{\bf P}-\frac{\bf q}{2})  
=  
\frac{1}{\pi^2}\int_{0}^{\infty} R^2\, dR \, j_{0}(qR)  
\Theta\left [ \hat k(R) - P \right ]  
\end{equation}  
with $\hat k(R) = \left [ 3\pi^2 \, \rho(R) \right ]^{1/3}$.  
  
The model equation for the $t$ matrix in terms of a reference {\it KN}   
potential in the  $K^{+}$N C.M. reference frame takes the   
form of the Lippmann-Schwinger type (c.f. Eq. \ref{kleing}), i.e.  
\begin{equation}  
\label{tmatrix}  
t_{KN}({\bf p}',{\bf p}; {s}) =  V_{KN}({\bf p}',{\bf p})  
+ 2\mu\int  \frac{d^3k}{(2\pi)^3}  
\frac{V_{KN}({\bf p}',{\bf k}) \; t_{KN}({\bf k},{\bf p}; {s})}  
{ k_\circ^2 + i\varepsilon - k^2} .  
\end{equation}  
Here, the energy invariant ${s}$ and associated on-shell momentum   
$k_\circ$ are determined from  
$s=(m_K + \epsilon_K + m_N + \bar\epsilon_N)^2 - {\bf  Q}^2$,  
where $E$ is the kinetic energy in the {\it KA} C.M. frame,   
$\bar\epsilon$ an average binding energy of the target nucleons  
and ${\bf Q}$ the total pair momentum.   
The potential $V_{KN}$ is constructed following the inversion procedure  
described in the previous Section.   
The calculation of the $t$ matrix on and off shell at various energies  
follows standard numerical procedures.   
In the boost of the $t$ matrix from the C.M. to the laboratory reference frame  
we have included the corresponding Jacobian (or M\o ller factor) \cite{Are02}.  
  
Although full-folding OMP were   
developed in the eighties for pion as well as nucleon scattering,   
most $K^{+}$-nucleus scattering analyses continue being   
made within an on-shell $t\rho$ approximation.   
We select and discuss three of these factorized forms in this study.  
  
\paragraph{Off-shell $t\rho$.}  
A first reduction to a $t\rho$ form emerges  
after setting ${\bf P}=0$ in the $t$ matrix in Eq. (\ref{FF}), thus allowing  
the integration of the mixed density over the momentum ${\bf P}$.  
Hence,  
\begin{equation}  
\label{trho}  
U({\bf k}^\prime, {\bf k}; E) =   
\sum_{N=p,n}   
\rho_{N}({\bf q})\;t_{NK^+} ({\bf k_r}',{\bf k_r};{\bf K};{s} )\;,  
\end{equation}  
where $\rho_N({\bf q})$ represents nuclear density in momentum space.  
In this factorized form the relative momenta   
${\bf k_r}$ and ${\bf k_r}'$ lie generally off shell, as no constraints   
on ${\bf k}$ nor ${\bf k}^\prime$ are in place.   
This reduction is referred as off-shell $t\rho$ approximation  
and has been extensively applied in nucleon-nucleus scattering.  
An additional step further can be taken to force the $t$ matrix on shell.  
Quite generally, features at the $t$-matrix level dictated by four independent   
variables (two magnitudes, angle and energy) are specified by two   
of its arguments, one angle and one energy.  
We have found that on-shell $t\rho$ results for $K^{+}A$ scattering depend,  
albeit moderately, on the prescription used and we focus on two of them.  
  
\paragraph{On-shell $t\rho$ of the $s-$type.}  
This is the usual form of the on-shell $t\rho$ approximation  
and has been applied extensively in hadron-nucleus collisions.  
We have named it of the $s-$type since it privileges the energy argument  
in the $t$ matrix.  
Basically, the energy $\sqrt{s}$ of the $K^{+}N$ pair is determined   
in the Breit frame with the subsequent determination, on-shell,  
of the relative momenta. Details can be found in Ref. \cite{Pae81}.  
  
\paragraph{On-shell $t\rho$ of the $k-$type.}  
An alternative prescription, which we refer as   
of the $k-$type, emerges naturally after considering a series expansion   
of $U({\bf k},{\bf k'})$ in terms of the magnitudes $k$ and $k'$ around   
the on-shell momentum $k_{A}$ in the projectile-nucleus C.M.  
Then, to lowest order we get  
\begin{equation}  
U({\bf k'},{\bf k}) \approx U(k_{A}{\bf \hat k'}, k_{A}{\bf \hat k}) \;.  
\end{equation}  
As a result, the two relative momenta in the $t$ matrix   
(c.f. Eq. (\ref{rel_mtum})) become equal in magnitude.  
The pair energy $\sqrt{s}$ is obtained on-shell from these  
relative momenta.

  
\section{Applications and results}  
  
We focus our applications on differential and total cross sections  
at kaon momenta in the range 400-1000 MeV/$c$ considering  
$^{6}$Li, $^{12}$C, $^{28}$Si and $^{40}$Ca targets.  
The ground-state densities of the first three targets  
were obtained from the nuclear charge density fit to  
electron scattering \cite{Li71a,Sic70,Li71b}.   
The point densities were obtained by unfolding the electromagnetic   
size of the proton from the charge density.   
In these cases we assume neutron densities equal to the proton densities.  
In the case of $^{40}$Ca we have used the densities    
from Ref. \cite{Neg70}.  
  
The scattering is analyzed within the full-folding OMP    
and comparisons are made with off- and on-shell $t\rho$ approximations.  
Thus, we include in the best possible way the off-shell effects in the  
effective interaction and switch them partially or fully off in the simpler  
$t\rho$  OMP.  
  
The {\it KA} optical potentials are calculated in momentum  
space following Ref. \cite{Are02}.  
The {\it KA} $S$ matrix and derived quantities linked with observable   
are obtained  by solving an OMP Lippmann-Schwinger equation for   
any of the specified nonlocal potentials.

\begin{figure}[!hbt]  
\includegraphics[scale=0.70,angle=00]{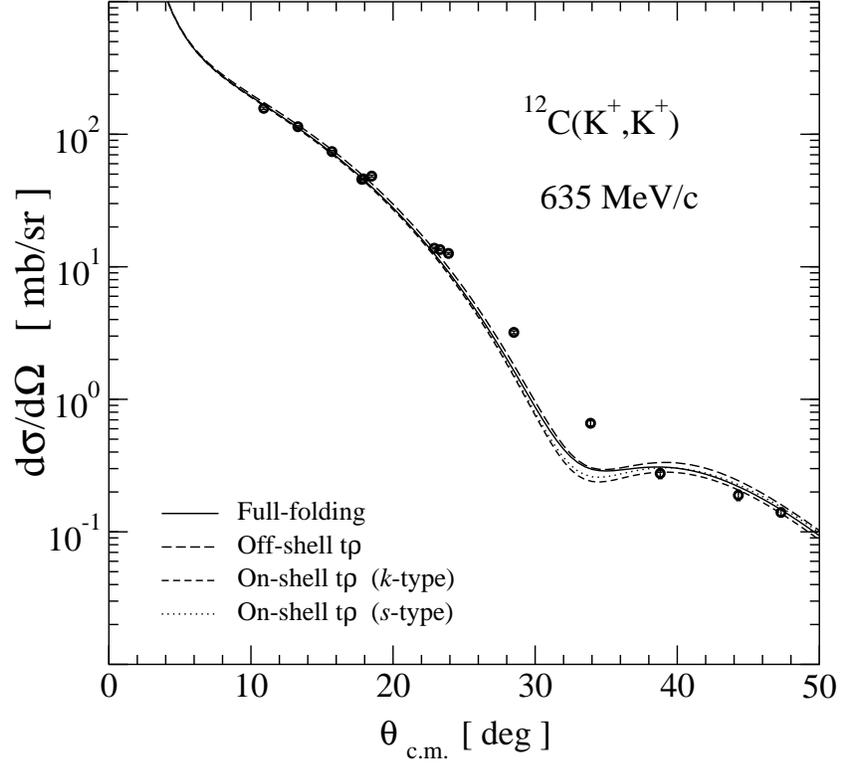}  
\vspace{-5mm}  
\caption{{\protect\small  
\label{C635}  
        Differential cross section for $K^+$+$^{12}$C elastic scattering  
        at P$_\textrm{Lab}$=635 MeV/$c$.  
        The solid curve represents the full-folding results,  
        whereas the long dashed curves corresponds to the off-shell  
        $t\rho$ results. The on-shell $t\rho$ results of the $k$-type  
        and $s$-type are shown with short-dashed and dotted curves,  
        respectively. The data are from Ref. \cite{Chr96}.  
        }}  
\end{figure}

In Fig. \ref{C635} we show the  calculated differential  
cross section for $K^{+}$+$^{12}$C scattering at beam momentum  635 MeV/$c$  
with  data \cite{Chr96}.  
The solid curves  are the full-folding results,  
whereas the long-dashed curves are the 
off-shell $t\rho$ results.  
The on-shell $t\rho$ results of the 
$s-$type and 
$k-$type   
are  shown as short-dashed and dotted curves, respectively.   
Although differences  exist among all four results, the differences are quite  
small.  
The differences among all 
$t\rho$ are a measure  
of the off-shell contributions.  
The 
off-shell $t\rho$  result show a uniform  
shift upward when compared with the full-folding results.  
More obvious, but still marginal, are  differences among the   
results for scattering angles above  $30$\,deg.  
  
\begin{figure}[!hbt]  
\includegraphics[scale=0.65,angle=-90]{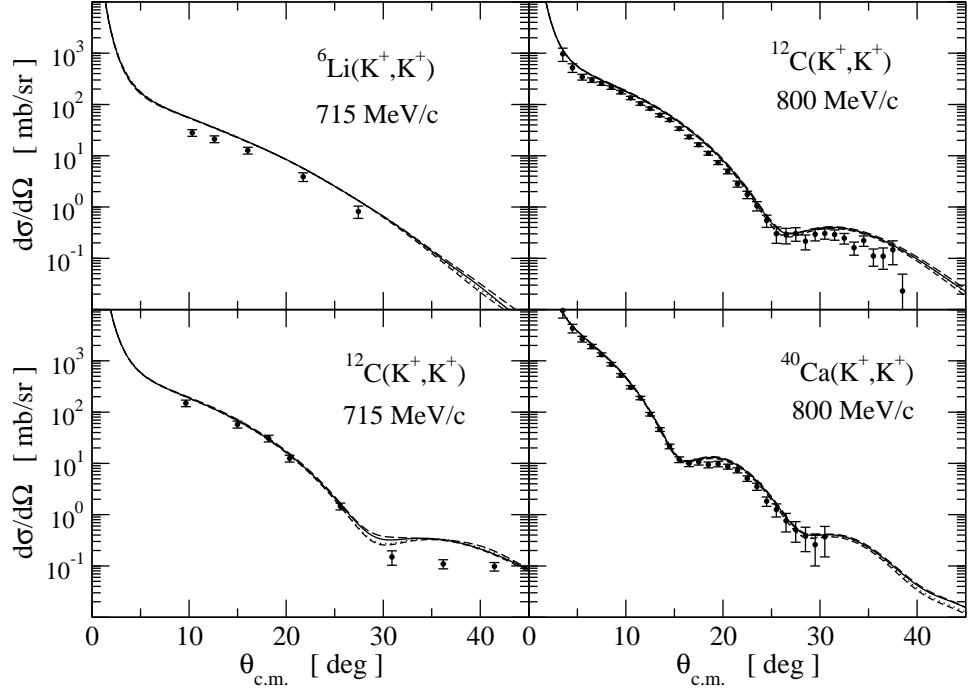}  
\vspace{-5mm}  
\caption{{\protect\small  
\label{FourX}  
        Differential cross section for elastic scattering of $K^{+}$ from   
        $^{6}$Li (upper-left frame),   
        $^{12}$C,  
        $^{40}$Ca (lower right frame) at beam momenta of   
        715 MeV/$c$ (left frames) and  
        800 MeV/$c$ (right frames).  
        The curve patterns are the same as in Fig. \ref{C635}.  
        The data at 715 MeV/$c$ and 800 MeV/$c$ are from Refs. \cite{Mic96}   
        and \cite{Marl82}, respectively.  
        }}  
\end{figure}

Similar applications are shown in Fig. \ref{FourX}, where we present  
 the differential cross section  for scattering  
from $^{6}$Li, $^{12}$C and $^{40}$Ca  
at beam momenta  715 MeV/$c$ (left frames), and 800 MeV/$c$ (right frames).  
The data are taken from \cite{Mic96,Marl82} and the curve  
textures follows the  convention of Fig. \ref{C635}.  
Here again we evidence moderate differences among all four approaches,  
being visible, at best, for angles above $25$\, deg.  
The comparison with data shows for $^{6}$Li (upper left frame)   
an overestimation of the theory with respect to the data  
by a factor of $\sim 1.8$  around 10 degrees.  
We favour to interpret this discrepancy being caused by uncertainties in   
the data normalization.   
The work by Chen {\em et al.} \cite{Che99}  
shows that they had a similar problem with $^{6}$Li.  
In their study they include a phenomenological second order  
potential proportional to a power of the nuclear density.  
They fit the complex strength and power of the density to the data,  
obtaining results in close resemblance to ours.  
Overall, the results for $^{12}$C and $^{40}$Ca shown in Fig. \ref{FourX}  
are in good accord with the data.  
In the case of $^{12}$C at 715 MeV/$c$ (lower left frame) some  
differences between  theory and data, at angles above $30$\, deg.  
are there. The full-folding and any of the $t\rho$ approaches  
are remarkably similar for differential cross sections.  
  
Total cross sections for $K^{+}$-nucleus  
have been extracted from transmission experiments \cite{Kra92,Fri97,Fri97b}.  
Such data are complementary to the differential cross section data and   
exhibit often larger differences among the full-folding and $t\rho$ results,  
even though the same {\it KN} effective interaction is used.  
Before jumping to fast conclusions about the effective interaction or  
the quality of any of the theoretical models, it is important to remember that  
the transmission total cross sections $\sigma_T$ have their own  model   
dependence built into  data. 
This has been discussed in some detail elsewhere \cite{Kau89,Ari93}.  
Using  $\sigma_T(\Omega)$ as the experimentally measured  
transmission cross section, subtending a solid angle $\Omega$ from the target  
along the beam axis, then the total cross section $\sigma_T$ is given by  
\begin{equation}  
\label{sigma_T}  
\sigma_T = \lim_{\Omega\rightarrow 0}   
\left [ \sigma_{T}(\Omega) - \sigma_{C}(\Omega_{>}) - \sigma_{CN}(\Omega_{>})   
\right] + \sigma_{N}(\Omega_{<}) + \sigma_{I}(\Omega_{<}) \;.  
\end{equation}  
Here $\Omega_{>}$ and $\Omega_{<}$ refers to the integrated cross section  
outside  and inside  the solid angle $\Omega$.  
Furthermore, we use the following  nomenclature for particular cross sections:  
 $\sigma_C(\Omega_{>})$ for the point charge Coulomb cross section,  
$\sigma_{CN}(\Omega_{>})$ for the Coulomb and nuclear interaction  interference term,  
$\sigma_N(\Omega_{<})$ for the nuclear cross section  from the nuclear   
interaction  
and $\sigma_I(\Omega_{<})$ arising from inelasticities.  
In the limit  $\Omega\rightarrow 0$ the last two terms vanish.  
However, $\sigma_{CN}$ requires very accurate results for the nuclear plus   
Coulomb interaction amplitude.   
This requires knowledge and availability  
of a high quality optical model in the first place, be as it may be,    
this introduces a model dependence of $\sigma_{T}$ which is beyond our  
judgment and puts limits on our conclusions.  
Nevertheless, we have calculated the total cross sections with all four    
optical models discussed here and compare the results with data reductions 
presented in Ref. \cite{Fri97} by Friedman {\em et al.}, and Ref. \cite{Fri97b}
by Friedman, Gal and Mar$\check{\rm e}$s.  
The difference between the data reported in these two references lies in the
way an optical potential, in a $t\rho(r)$ form, is constructed to extract
the total cross sections from transition experiments.
Whereas in Ref. \cite{Fri97} the $t\rho$ form is based on a density independent
$t$-matrix strength, in Ref. \cite{Fri97b} the imaginary part of the strength 
exhibits a parametric density dependence adjusted to yield, self-consistently,
the total cross sections. Thus, the data reported in the second reference
is consistent within an empirical medium dependence 
(c.f. Eq. (5) of Ref. \cite{Fri97b}) of the $t$ matrix and reflects, 
to some extent, the model dependence of their reported measurements.
  
\begin{figure}[!hbt]  
\includegraphics[scale=0.70,angle=-00]{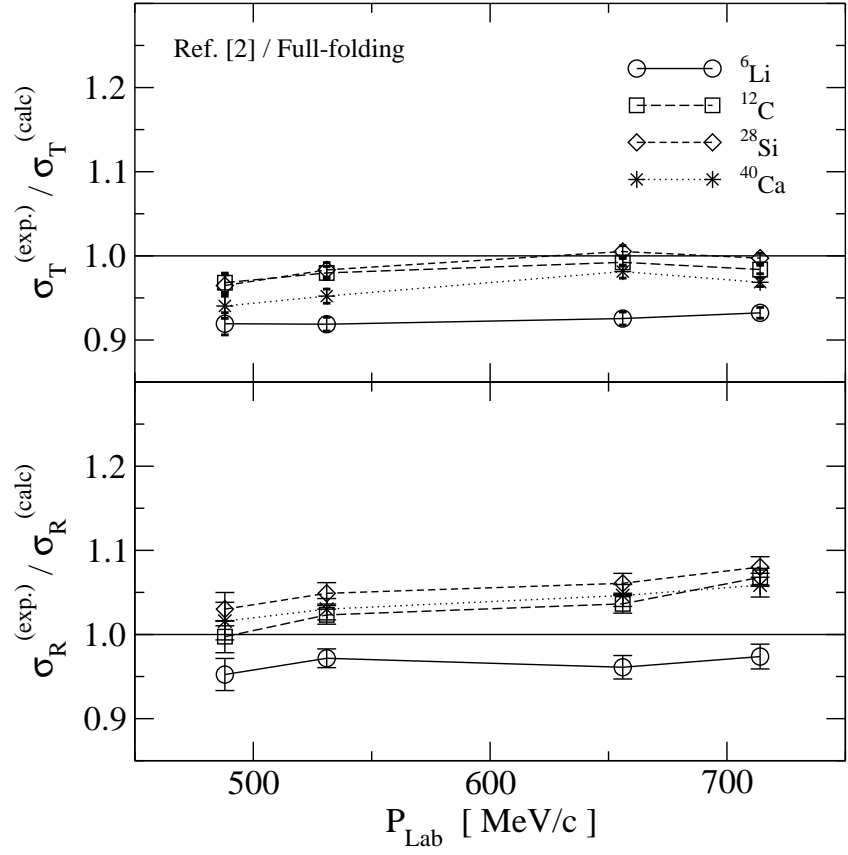}  
\vspace{-5mm}  
\caption{{\protect\small  
\label{Ratios-ff}  
        Experimental-to-calculated ratios for the total   
        ($\sigma_T$) and reaction ($\sigma_R$) cross sections   
        for $K^{+}$ elastic scattering from $^{6}$Li, $^{12}$C,   
        $^{26}$Si and $^{40}$Ca, at 488, 531, 656 and 714 MeV/$c$.  
        Results based on the full-folding approach.  
        Connecting lines have been drawn to guide the eye.  
	The data are taken from Ref. \cite{Fri97}.
        }}  
\end{figure}

\begin{figure}[!hbt]  
\includegraphics[scale=0.70,angle=-00]{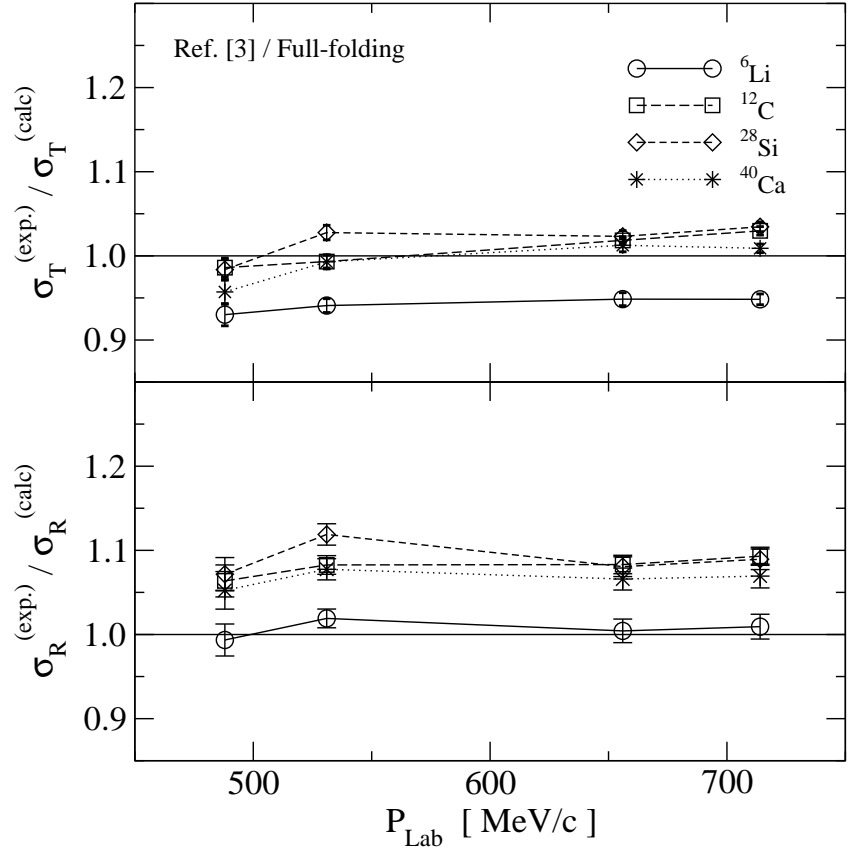}  
\vspace{-5mm}  
\caption{{\protect\small  
\label{Ratios-ff-b}  
        The same as Fig. \ref{Ratios-ff}, but with 
	the data taken from Ref. \cite{Fri97b}.
        }}  
\end{figure}

In Figs. \ref{Ratios-ff} and \ref{Ratios-ff-b}  we present the ratios 
experiment/calculated   
of the reaction cross sections 
$\sigma_R\mbox{(\textrm{Exp.})}/\sigma_R\mbox{(\textrm{Calc.})}$   
and  the total cross sections 
$\sigma_T\mbox{(\textrm{Exp.})}/\sigma_T\mbox{(\textrm{Calc.})}$   
for  four target nuclei at four projectile momenta.  
We notice that all ratios are nearly constant as function of projectile   
momentum,   
whereas only the $^{6}$Li results lie somewhat below the other three cases.  
Quite similar results are obtained considering the other three forms
of the $t\rho$ model.  
When comparing Figs. \ref{Ratios-ff} and \ref{Ratios-ff-b} we observe
a clear shift in the reaction cross section of the latter with respect to
the former. 
This shift is consistent with the rescaling of the imaginary
part of the strength of the $t$ matrix used in the construction of the
optical potential \cite{Fri97b}. 
The question is, therefore, whether this prescription to incorporate
medium corrections effectively accounts for
genuine medium effects in the form of short range correlations,
Fermi motion and their implied non local effects in the $K^{+}$-nucleus
coupling. 
An assessment of these issues remains to be seen.
  
\begin{figure}[ht]  
\includegraphics[scale=0.55,angle=-90]{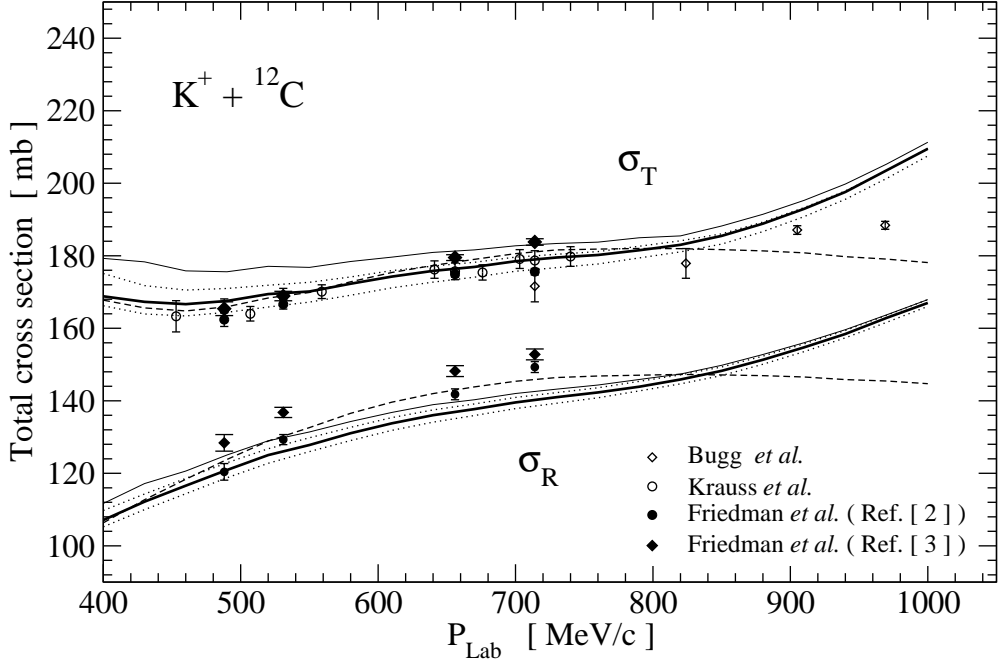}  
\vspace{-5mm}  
\caption{{\protect\small  
\label{Total_XS}  
        Measured and calculated total cross sections as functions of the  
        beam momentum for $K^{+}$ + $^{12}$C scattering.  
        The thick curves represent full-folding results;  
        the dotted and thinner solid curves correspond to on- and  
        off-shell $t\rho$ results, respectively.  
        The dashed curves represent full-folding results with  
        the separable strength of the elemental $K^{+}N$ potential suppressed.  
        The data are from Refs. \cite{Fri97,Fri97b,Kra92,Bug68}  
        }}  
\end{figure}

The features observed above can also be seen in the Table \ref{xsection},  
where we present the measured and calculated cross  
sections at four momenta for the selected targets,  
from calculations based on the four approaches discussed here.  
For instance, the results shown in Fig. \ref{Ratios-ff}   
correspond to the ratios between the first two blocks of this Table.  
When comparing the full-folding cross sections with the on-shell $t\rho$   
results, we observe that the   
former lies systematically above the 
$k-$type, but below the  
$s-$type.  
These differences may be used to estimate the off-shell sensitivity,   
which we estimate  $\pm$3\% for the worst case.  
The 
off-shell $t\rho$  
results is always  above the other three results and  its difference to   
the data is the largest.  
These features become more evident in Fig. \ref{Total_XS},   
where we present the measured and calculated   
reaction $\sigma_R$ and total $\sigma_T$ cross sections   
for $^{12}$C as function of the kaon momentum,  
in the range 400-1000 MeV/$c$.  
The data from 
Bugg {\em et al.} \cite{Bug68} and 
Krauss {\em et al.} \cite{Kra92} are shown with open diamonds and circles,
respectively. The data form Friedman {\em et al.} \cite{Fri97}, and
Friedman, Gal and Mar$\check{\rm e}$s \cite{Fri97b} are shown
with black circles and diamonds, respectively.
Here, the thicker solid curves represent the full-folding results,  
whereas the dotted ones are based on the 
on-shell $t\rho$ approaches.  
The 
off-shell $t\rho$ results  are shown with the thinner   
solid curves.  
Finally, we  consider of interest to present full-folding results   
when only the reference inversion potentials are used in the   
{\it KN} effective interaction, being the separable contribution 
completely suppressed.  
These results for $\sigma_T$ and $\sigma_R$ are shown with dashed curves.  
  
The full-folding and $t\rho$ approaches give an overall consistent agreement  
with the measured total cross sections up to 900 MeV/$c$, above which  
they depart from the data.  
Notice that a nearly full agreement -within error bars- is achieved with  
the data of Krauss {\em et al.} \cite{Kra92}.  
Furthermore, the 
$s-$type $t\rho$  $\sigma_T$ results   
(upper dotted curves) are in less good agreement with the data  
in comparison with the 
$k-$type $t\rho$ and full-folding approach,   
particularly below 500 MeV/$c$.   
This illustrates the relevance of the target nucleon Fermi motion   
in the 
off-shell $t\rho$  
results below $\sim$700 MeV/$c$,  shown as thin solid curves,   
which lies distinctively above the data.  
A drifting apart of  all curves at the lower  
momenta supports the proper inclusion of Fermi motion in the   
treatment of the effective interaction.

A closer scrutiny of the gradual departure of the calculated total   
cross section relative to the data, above 900 MeV/$c$,  
would require the study of possible uncertainties in the elemental   
$K^{+}N$ amplitude and to assess their impact on total cross sections.  
These considerations go beyond the focus of the present work.  
Incidentally, the results where separable strength of $V_{KN}$ is   
suppressed (dashed curves) indicate that, despite marginal differences in   
the description of the real phase shifts,   
the absorptive component becomes important  in the asymptotic behavior  
of the cross sections.  
It is in this high-energy regime where  
the single- and continuous-energy solutions exhibit sizeable differences.  
  
The sensitivity of $\sigma_T$ to the alternative approaches considered here   
is somewhat diminished in the context of the reaction cross section,  
where all curves stay much closer to each other.  
An interesting feature which emerges after comparing the calculated total 
and reaction cross sections is their nearly constant difference 
above 600 MeV/$c$.
In the particular case of $^{12}$C we observe
\begin{equation}
\sigma_T \approx \sigma_R + \textrm{39 [mb]}.  
\end{equation}
A similar behavior is exhibited by the other targets,   
as inferred from Table \ref{xsection}.  

\begin{figure}[ht]  
\includegraphics[scale=0.55,angle=-90]{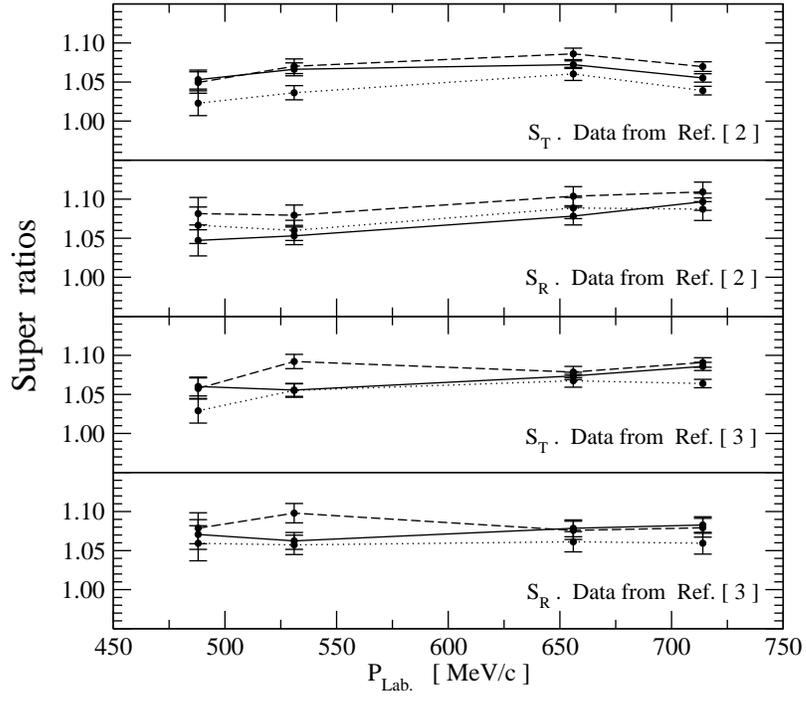}  
\vspace{-5mm}  
\caption{{\protect\small  
\label{SSratios}
        Super ratios for total and reaction cross sections
	based on data from Refs. \cite{Fri97, Fri97b} and the 
	full-folding approach.
	The solid, dashed and dotted curves represent results 
	for $^{12}$C, $^{28}$Si and $^{40}$Ca, respectively.
        }}  
\end{figure}

The study of total cross sections for $N=Z$ nuclei have also
been of some interest as a means to gauge the role of medium effects
in the propagation of kaons through the nucleus.
Weiss and collaborators \cite{Wei94} found that the ratio 
$\sigma_T/A$ for $^{6}$Li and deuterium are nearly the same, suggesting 
that multiple steps contributions are rather weak in these light targets.
Such is not the case for the heavier targets.
In order to quantify this feature, Friedman {\em et al.} have introduced
the super ratios, {\em i.e.} the ratio 
$\sigma_\textrm{Exp.}(A)\sigma_\textrm{Calc.}(^6\textrm{Li})/
 \sigma_\textrm{Calc.}(A)\sigma_\textrm{Exp.}(^6\textrm{Li})$.
Although it is correct that this quantity would diminish normalization
uncertainties, its departure from one may not only indicate medium effects
but also the level of disagreement between theory and experiment.
Indeed, their reported values for each target exhibit distinctive curves
as function of the momentum, with values ranging between 1.15 and 1.25.
Although limited by the fact that the optical model used to extract the data
in Refs. \cite{Fri97,Fri97b} differs from the full-folding model used here, 
we have also calculated the super ratios using the results in Table I.
In Fig. \ref{SSratios} we plot the total ($S_T$) and reaction ($S_R$) 
super ratios considering the data from 
Ref. \cite{Fri97} (upper two frames)
and Ref. \cite{Fri97b} (lower two frames), against the full-folding results.
Notice that all super ratios are nearly constant as functions of the 
momentum, with variations between 1.0 and 1.1, consistent with the level
of agreement shown in Figs. \ref{Ratios-ff} and \ref{Ratios-ff-b}.
Nonetheless, definite analyses of these super ratios requires
the use of full-folding $K^{+}A$ optical potentials to extract the 
cross section data from transmission experiments, 
an endeavor beyond the scope this work.

  
\section{Conclusions}  
  
We have studied $K^{+}$-nucleus elastic scattering from light nuclei  
in the momentum range 400-1000 MeV/$c$ within the full-folding optical   
model potential framework.   
To this purpose we have used the $t$ matrix based on a $K^{+}N$   
potential model with absolute match of the phase shift analyses reported  
by the GWU group.  
The emphasis here has been placed on a strict connection between the   
bare $K^{+}N$ potential -consistent with the current phase shift analysis-   
and the $K^{+}$-nucleus scattering process.  
This feature is achieved by adding a separable term to a local reference  
potential obtained within the Gel'fand-Levitan-Marchenko quantum inversion.  
The $t$ matrix, based on this bare potential model, is convoluted with the   
nuclear mixed density leading to a nonlocal $K^{+}A$ optical potential.  
The scattering observable were compared with those obtained within   
off-shell and two alternative versions of on-shell $t\rho$ approximations,  
which we have named of the $s-$ and $k-$type, respectively.  
Considering the differential cross-sections,   
we observe moderate differences among the calculated results from  
all four approaches.  
However, reaction and total cross-sections for transmission experiments  
show a clear sensitivity to the way the Fermi motion is treated, being this  
more notorious at the lower momenta, i.e. P$_\textrm{Lab}\lesssim$ 600 MeV/$c$. 
Furthermore, an excellent account of the total cross sections reported by  
Krauss {\em et al.} \cite{Kra92} is provided by the full-folding approach   
for momenta between 
450 MeV/$c$ $\lesssim$ P$_\textrm{Lab} \lesssim$ 750 MeV/$c$.  
These results demonstrate that definite  
conclusions about the ability of any microscopic approach to describe   
total cross section data must include   
the genuine off-shell behavior of the effective interaction and its  
energy dependence.   
These conventional medium effects become essential before any   
conclusive assessment about the manifestation of subhadronic degrees of 
freedom in {\it KA} collisions, particularly a possible manifestation of 
$\Theta^{+}$(1540) in the collision of $K^{+}$ with nuclei \cite{Gal05}.  
  
\begin{acknowledgments}  
H. F. A. acknowledges the Nuclear Theory Group of the   
University of Hamburg for its kind hospitality during his visit.  
Partial funding for this work was provided by FONDECYT under   
Grant No 1040938.  
\end{acknowledgments}

\newpage  
 \begin{table*}  
 \caption{\label{xsection}  
  Experimental and calculated reaction and total cross sections   
 (in mb) for $K^{+}$-nucleus scattering at the specified momenta.  
 The data in the first and second block are from Refs. \cite{Fri97,Fri97b}
and their corresponding errors are quoted between parentheses.}  
 \begin{tabular} {lc llll c llll }  
 \hline  
 &          & \multicolumn{4}{c}{Reaction} &\rule[0pt]{11pt}{0pt}&\multicolumn{4}{c}{Total} \\  
Source & P$_\textrm{Lab}$ \\  
 & [MeV/$c$]  & $^{6}$Li & $^{12}$C & $^{28}$Si & $^{40}$Ca && $^{6}$Li & $^{12}$C & $^{28}$Si & $^{40}$Ca \\  
 \hline  
 Data \cite{Fri97}  
    &  488 &  65.0(1.3) & 120.4(2.3) & 265.5(5.1) & 349.9(7.7) &&  
              76.6(1.1) & 162.4(1.9) & 366.5(4.8) & 494.4(7.7) \\  
    &  531 &  69.8(0.8) & 129.3(1.4) & 280.4(3.4) & 367.1(4.5) &&  
              78.8(0.7) & 166.6(1.3) & 374.8(3.3) & 500.2(4.4) \\  
    &  656 &  75.6(1.1) & 141.8(1.5) & 306.1(3.4) & 401.1(5.0) &&  
              84.3(0.7) & 174.9(0.8) & 396.1(2.7) & 531.9(4.2) \\  
    &  714 &  79.3(1.2) & 149.3(1.5) & 317.5(3.6) & 412.9(5.5) &&  
              87.0(0.6) & 175.6(0.9) & 396.5(2.3) & 528.4(2.8) \\  
 \hline  
 Data \cite{Fri97b}  
    &  488 &  67.8(1.3) & 128.4(2.3) & 276.2(5.1) & 362.5(7.7) &&  
              77.5(1.1) & 165.4(1.9) & 373.7(4.8) & 503.2(7.7) \\  
    &  531 &  73.2(0.8) & 136.8(1.4) & 299.1(3.4) & 384.0(4.5) &&  
              80.7(0.7) & 168.9(1.3) & 391.7(3.3) & 521.6(4.4) \\  
    &  656 &  79.0(1.1) & 148.2(1.5) & 311.8(3.4) & 408.6(5.0) &&  
              86.4(0.7) & 179.5(0.8) & 403.2(2.7) & 548.8(4.2) \\  
    &  714 &  82.2(1.2) & 152.8(1.5) & 320.2(3.6) & 417.1(5.5) &&  
              88.5(0.6) & 183.8(0.9) & 411.3(2.3) & 550.4(2.8) \\  
\hline  
Full-folding  
  &  488 &  68.2 & 120.7 & 257.7 & 344.4 &&  83.3 & 167.7 & 379.9 & 525.8 \\  
  &  531 &  71.8 & 126.4 & 267.3 & 356.4 &&  85.8 & 170.1 & 381.2 & 525.4 \\  
  &  656 &  78.7 & 136.8 & 288.5 & 383.3 &&  91.1 & 176.3 & 394.1 & 542.0 \\  
  &  714 &  81.4 & 139.8 & 293.9 & 390.0 &&  93.3 & 178.5 & 397.6 & 545.5 \\  
 \hline  
Off-shell $t\rho$  
  &  488 &  70.1 & 124.8 & 265.1 & 353.7 &&  86.3 & 175.8 & 396.5 & 548.4 \\  
  &  531 &  73.5 & 129.9 & 274.0 & 364.7 &&  88.3 & 176.9 & 395.7 & 545.3 \\  
  &  656 &  79.9 & 139.6 & 293.0 & 388.7 &&  92.9 & 181.2 & 403.2 & 554.0 \\  
  &  714 &  83.0 & 141.9 & 298.2 & 395.2 &&  94.7 & 182.5 & 405.8 & 556.3 \\  
 \hline  
$t\rho$ $k-$type  
  &  488 &  67.7 & 118.5 & 252.4 & 336.8 &&  82.7 & 164.4 & 369.3 & 508.5 \\  
  &  531 &  71.2 & 124.1 & 262.7 & 349.8 &&  85.0 & 166.4 & 371.6 & 510.3 \\  
  &  656 &  78.3 & 135.0 & 284.9 & 378.4 &&  90.5 & 173.4 & 386.8 & 530.8 \\  
  &  714 &  80.3 & 138.6 & 291.3 & 386.2 &&  92.2 & 176.3 & 391.5 & 535.9 \\  
 \hline  
$t\rho$ $s-$type  
  &  488 &  69.2 & 122.6 & 260.4 & 347.4 &&  84.9 & 170.9 & 382.9 & 527.6 \\  
  &  531 &  72.7 & 128.0 & 270.2 & 359.7 &&  87.0 & 172.2 & 383.6 & 527.0 \\  
  &  656 &  79.5 & 138.3 & 291.1 & 386.6 &&  92.0 & 177.6 & 395.9 & 543.5 \\  
  &  714 &  81.4 & 141.6 & 297.1 & 393.8 &&  93.4 & 180.1 & 399.7 & 547.4 \\  
 \hline  
 \end{tabular}  
 \end{table*}  
\newpage  
  
\end{document}